\documentclass [reprint,prd,aps,twocolumn]{revtex4-1}

\usepackage{graphicx}
\usepackage{color}
\usepackage{amsfonts}
\usepackage{amsmath}
\usepackage{ulem}

\begin{document}
\title{Remarks on the phenomenological Tsallis distributions and their link with the Tsallis statistics}

\author{A.S.~Parvan}
\email{parvan@theor.jinr.ru, parvan@theory.nipne.ro}
\affiliation{\textit{Bogoliubov Laboratory of Theoretical Physics, Joint Institute for Nuclear Research, Dubna, Russia} \\\textit{Department of Theoretical Physics, Horia Hulubei National Institute of Physics and Nuclear Engineering, Bucharest-Magurele, Romania}\\
\textit{Institute of Applied Physics, Moldova Academy of Sciences, Chisinau, Republic of Moldova}}
\author{T.~Bhattacharyya}
\email{bhattacharyya@theor.jinr.ru}
\affiliation{\textit{Bogoliubov Laboratory of Theoretical Physics, Joint Institute for Nuclear Research, Dubna, Russia}}

\begin{abstract}
From the Tsallis unnormalized (or Tsallis-2) statistical mechanical formulation, B\"{u}y\"{u}kkili\c{c} {\it et al.} [Phys. Lett. A 197, 209 (1995)] derived the expressions for the single-particle distribution functions (known as the phenomenological Tsallis distributions) for particles obeying the Maxwell-Boltzmann, Bose-Einstein and the Fermi-Dirac statistics using the factorization approximation. In spite of the fact that this paper was published long time ago, its results are still extensively used in many fields of physics, and it is considered that it was this paper that established the connection between the phenomenological Tsallis distributions and the Tsallis statistics. Here we show that this result is incorrect: the mistake lies in the fact that the probability distribution function was derived using the definition of the generalized expectation values (of the Tsallis-2 statistics), but the single-particle distribution function was calculated from this probability distribution using the standard definition of the expectation values of the Tsallis normalized (or Tsallis-1) statistics. Considering the definition of the expectation values which is consistent with the Tsallis-2 formulation, we have proved that the single-particle (classical and quantum) distribution functions in the factorization approximation differ from the phenomenological Tsallis distributions.
\end{abstract}


\maketitle

\section{Introduction}
The phenomenological single-particle Tsallis distributions for the classical and quantum statistics of particles introduced in Ref.~\cite{Buyukkilic93} (inspired by the Tsallis statistics~\cite{Tsal88}) and explicitly derived in the factorization approximation from the Tsallis statistics in Ref.~\cite{Buyukkilic} have gained much attention. These distributions are widely used in many fields of physics such as high energy  collisions~\cite{PHENIX1,CMS1,CMS2,ALICE_deuteron,Bediaga00,Beck00,Wilk09,Biro09,Cleymans12,Cleymans09,Cleymans12a,Wong15,Rybczynski14,Cleymans13,Azmi14,Marques13,Grigoryan17,Li14,Parvan17,Marques15,TsallisTaylor,TsallisRAA,BhattaCleMog,bcmmp,Shen18}, Bose-Einstein condensation~\cite{Miller06,Lawani08,Biswas08}, black-body radiation~\cite{Iwasaki12,Wang98}, neutron star~\cite{Menezes15,Megias15}, early universe~\cite{Pessah01}, superconductivity~\cite{Nunes02}, etc. They have the form
\begin{equation}\label{i}
  \langle n_{\mathbf{p}\sigma} \rangle = \left[1-(1-q)\frac{\varepsilon_{\mathbf{p}}-\mu}{T}\right]^{\frac{1}{1-q}}
\end{equation}
for the Maxwell-Boltzmann statistics of particles and
\begin{equation}\label{ii}
  \langle n_{\mathbf{p}\sigma} \rangle = \frac{1}{\left[1-(1-q)\frac{\varepsilon_{\mathbf{p}}-\mu}{T}\right]^{\frac{1}{q-1}}\pm 1}
\end{equation}
for the Bose-Einstein (minus sign) and Fermi-Dirac (plus sign) statistics of particles~\cite{Buyukkilic93,Buyukkilic}, where $\varepsilon_{\mathbf{p}}$ is a single-particle energy. These single-particle (classical and quantum) distribution functions are popularly known as the Tsallis distributions and in literature they are described as belonging to the Tsallis nonextensive statistics~\cite{Tsal88} which introduces a generalized definition of entropy. There are several schemes for the Tsallis nonextensive statistics~\cite{Tsal98}. The Tsallis normalized (Tsallis-1) and Tsallis unnormalized (Tsallis-2) statistics are two examples of them. The phenomenological Tsallis distributions (\ref{i}) and (\ref{ii}) were obtained in Ref.~\cite{Buyukkilic93} using the method of the maximization of the generalized entropy of the ideal gas. Refs.~\cite{Cleymans12,Cleymans12a} derived the single-particle Tsallis distributions by using the same method. The general problems of this method in the case of the Tsallis statistics were discussed in Ref.~\cite{Parvan2017a}. The single-particle Maxwell-Boltzmann distribution (\ref{i}) was also shown to be the stationary state solution of the generalized Boltzmann kinetic equation~\cite{Lavagno,Biro05,Alberico09,Osada08} and nonlinear Fokker-Planck equation~\cite{Tsallis96,Lavagno02a}. Nonextensive relativistic kinetic theory with quantum statistical effects was elaborated in Ref.~\cite{Mitra18}.

But, such approaches lack the connection between the single-particle distributions (\ref{i}) and (\ref{ii}) and the probability distribution of microstates of the Tsallis statistics~\cite{Tsal88,Tsal98}. An attempt to establish this connection was made in the paper of F.~B\"{u}y\"{u}kkili\c{c} {\it et al.}~\cite{Buyukkilic} in which the closed form analytical formulas (\ref{i}) and (\ref{ii}) for the single-particle distribution functions of the Bose-Einstein, Fermi-Dirac and Maxwell-Boltzmann statistics of particles were directly derived from the probability distribution of microstates of the Tsallis unnormalized statistics by using the factorization approximation. In Ref.~\cite{Buyukkilic}, it was demonstrated that the phenomenological distribution functions (\ref{i}) and (\ref{ii}) correspond to the Tsallis unnormalized (also known as Tsallis-2) statistics~\cite{Tsal88,Tsal98} in the factorization approximation. However, we observed an inconsistency in the definition of the average value used in calculating the average number of particles in a microstate in Ref. \cite{Buyukkilic}. To be more specific, Eqs.~(35) and (64) of Ref.~\cite{Buyukkilic} lack power $q$ of the probability distribution of microstates, which eventually would have been correct had one been dealing with the Tsallis normalized (or Tsallis-1) statistics~\cite{Tsal88,Tsal98} instead of the Tsallis-2 statistics. In this article, we present a consistent derivation of the single-particle distribution functions of the Bose-Einstein, Fermi-Dirac and Maxwell-Boltzmann statistics of particles in the framework of the Tsallis unnormalized statistics in the factorization approximation using the definition of the generalized statistical averages of the Tsallis-2 statistics. We show that the single-particle distribution functions of the Tsallis unnormalized statistics in the factorization approximation differ from the single-particle distributions (\ref{i}) and (\ref{ii}) obtained in Ref.~\cite{Buyukkilic}.

It is worthwhile to mention that the link of the single-particle distributions (\ref{i}) and (\ref{ii}) with the probability distribution of the microstates in the Tsallis statistics was also studied for the massless \cite{Parvan2017a} and massive \cite{Parvan19} particles without introducing the factorization approximation. In these papers, it was demonstrated that the phenomenological single-particle Tsallis distribution (\ref{i}) for the Maxwell-Boltzmann statistics of particles corresponds to the Tsallis unnormalized statistics in the zeroth term approximation. For the massive quantum particles, however, this observation might not be generalized.

After these introductory discussions, we now move towards demonstrating our main findings in the forthcoming sections. The paper is organized as follows. In the next section we discuss the general formalism of the Tsallis unnormalized statistics used in Ref.~\cite{Buyukkilic}. In Sec.~\ref{sec3} we derive the classical and quantum single-particle distribution functions in the Tsallis unnormalized statistics with and without the factorization approximation. In Sec.~\ref{sec4} we repeat and reproduce the calculations of Ref.~\cite{Buyukkilic} to find out the classical and quantum single-particle distribution functions and contrast them with those obtained in Sec.~\ref{sec3}. And lastly, we summarize and conclude in Sec.~\ref{sec5}.

\section{Tsallis unnormalized statistics}\label{sec2}
The authors of Ref.~\cite{Buyukkilic} use the Tsallis unnormalized (Tsallis-2) statistics~\cite{Tsal88} which is defined by the generalized entropy in terms of the probabilities $\{p_{i}\}$ of the microstates normalized to unity~\cite{Tsal88,Tsal98}
\begin{equation}\label{1}
    S =  \sum\limits_{i} \frac{p_{i}^{q}-p_{i}}{1-q}, \qquad  \sum\limits_{i} p_{i}=1
\end{equation}
and by the generalized expectation values~\cite{Tsal88,Tsal98}
\begin{equation}\label{2}
   \langle A \rangle = \sum\limits_{i} p_{i}^{q} A_{i},
\end{equation}
where $q\in\mathbb{R}$ is a real parameter taking values $0<q<\infty$. In the Gibbs limit $q\to 1$, the entropy (\ref{1}) recovers the Boltzmann-Gibbs-Shannon entropy, $S=-\sum_{i} p_{i} \ln p_{i}$, and the Tsallis-2 statistics is reduced to the Boltzmann-Gibbs one. It should be stressed that the Tsallis-2 statistics is unnormalized because the expectation values (\ref{2}) are not consistent with the norm equation of the probabilities of microstates  given in Eq.~(\ref{1}). The probabilities in the norm equation have a linear dependence, but in the definition of the expectation values they are in the power of $q$.

The thermodynamic potential of the grand canonical ensemble of the Tsallis-$2$ statistics can be written as~\cite{Parvan2017a}
\begin{equation}\label{3}
 \Omega = \langle H \rangle -TS-\mu \langle N \rangle = \sum\limits_{i}  p_{i}^{q} \left[E_{i}-\mu N_{i} + T \frac{p_{i}^{1-q}-1}{1-q}\right],
\end{equation}
where $\langle H \rangle=\sum_{i}  p_{i}^{q} E_{i}$ is the mean energy of the system and $\langle N \rangle=\sum_{i}  p_{i}^{q} N_{i}$ is the mean number of particles of the system.

The unknown probabilities $\{p_{i}\}$ of the Tsallis statistics are found in the point of equilibrium of the system from the second law of thermodynamics or Jaynes principle~\cite{Jaynes2} using the constrained local extrema of the thermodynamic potential~\cite{Parvan2015} by the method of the Lagrange multipliers (see, for example, Ref.~\cite{Krasnov}):
\begin{eqnarray}\label{4}
  \frac{\partial \Phi}{\partial p_{i}} &=& 0, \\ \label{5}
 \Phi &=& \Omega - \lambda \varphi, \\ \label{6}
  \varphi &=& \sum\limits_{i} p_{i} - 1 = 0,
\end{eqnarray}
where $\Phi$ is the Lagrange function and $\lambda$ is an arbitrary real constant. Then, we get~\cite{Tsal98,Parvan2017a}
\begin{eqnarray}\label{7}
p_{i} &=& \frac{1}{Z}  \left[1-(1-q) \frac{E_{i}-\mu N_{i}}{T} \right]^{\frac{1}{1-q}}, \\ \label{8}
    Z &=& \sum\limits_{i} \left[1-(1-q) \frac{E_{i}-\mu N_{i}}{T} \right]^{\frac{1}{1-q}},
\end{eqnarray}
where $Z\equiv [(1-(1-q)\lambda/T)/q]^{1/(1-q)}$ is the norm function related to the Lagrange multiplier $\lambda$, which is fixed by the norm equation of probabilities given in  Eq.~(\ref{1}) (see Refs.~\cite{Tsal98,Parvan2017a}). Thus the statistical averages (\ref{2}) for the Tsallis-$2$ statistics in the grand canonical ensemble can be rewritten in the general form as~\cite{Tsal98,ParvanBaldin}
\begin{equation}\label{9}
   \langle A \rangle = \frac{1}{Z^{q}}\sum\limits_{i} A_{i} \left[1-(1-q)\frac{E_{i}-\mu N_{i}}{T}\right]^{\frac{q}{1-q}},
\end{equation}
where the partition function $Z$ is calculated by Eq.~(\ref{8}). The general formulas for the Tsallis-2 statistics in the integral representation can be found in Ref.~\cite{Parvan19}.

\section{Single-particle distribution function of Tsallis unnormalized statistics}\label{sec3}
\subsection{Exact results}
Let us consider the ideal gas for the Tsallis-$2$ statistics in the grand canonical ensemble and calculate the mean occupation numbers with and without the factorization approximation. In the occupation number representation, the mean occupation number (obtainable from the definition in Eq. (\ref{9})) and the partition function (defined in Eq. (\ref{8})) for the ideal gas in the Tsallis-$2$ statistics in the grand canonical ensemble can be written as~\cite{Parvan2017a,Parvan19}
\begin{eqnarray}\label{10}
  \langle n_{\mathbf{p}\sigma} \rangle &=& \frac{1}{Z^{q}} \sum\limits_{\{n_{\mathbf{p}\sigma}\}} n_{\mathbf{p}\sigma} G\{n_{\mathbf{p}\sigma}\} \nonumber \\ &\times& \left[1-(1-q)\frac{\sum\limits_{\mathbf{p},\sigma} n_{\mathbf{p}\sigma} (\varepsilon_{\mathbf{p}}-\mu)}{T}\right]^{\frac{q}{1-q}}
\end{eqnarray}
and
\begin{equation}\label{11}
  Z=\sum\limits_{\{n_{\mathbf{p}\sigma}\}}  G\{n_{\mathbf{p}\sigma}\}   \left[1-(1-q)\frac{\sum\limits_{\mathbf{p},\sigma} n_{\mathbf{p}\sigma} (\varepsilon_{\mathbf{p}}-\mu)}{T}\right]^{\frac{1}{1-q}},
\end{equation}
where $\varepsilon_{\mathbf{p}}=\sqrt{\mathbf{p}^2+m^{2}}$ is the single-particle energy, $m$ is the mass of the particle, $n_{\mathbf{p}\sigma}=0,1,\ldots,K$ are the occupation numbers, $G\{n_{\mathbf{p}\sigma}\}=1$ for the Fermi-Dirac $(K=1)$ and Bose-Einstein $(K=\infty)$ statistics of particles, and $G\{n_{\mathbf{p}\sigma}\}=1/(\prod_{\mathbf{p},\sigma}n_{\mathbf{p}\sigma}!)$ for the Maxwell-Boltzmann $(K=\infty)$ statistics of particles.

The probability distribution (\ref{7}) in the occupation number representation is~\cite{Parvan19}
\begin{equation}\label{12}
 p\{n_{\mathbf{p}\sigma}\} = \frac{1}{Z} \left[1-(1-q)\frac{\sum\limits_{\mathbf{p},\sigma} n_{\mathbf{p}\sigma} (\varepsilon_{\mathbf{p}}-\mu)}{T}\right]^{\frac{1}{1-q}}.
\end{equation}
Then, the mean occupation numbers (\ref{10}) can be rewritten as
\begin{eqnarray}\label{13}
  \langle n_{\mathbf{p}\sigma} \rangle &=& \sum\limits_{\{n_{\mathbf{p}\sigma}\}} n_{\mathbf{p}\sigma} G\{n_{\mathbf{p}\sigma}\}  \left(p\{n_{\mathbf{p}\sigma}\}\right)^{q}  \\ \label{13a}  &=& \sum\limits_{n_{\mathbf{p}\sigma}=0}^{K} n_{\mathbf{p}\sigma} f(n_{\mathbf{p}\sigma})
\end{eqnarray}
and
\begin{eqnarray}\label{14}
   f(n_{\mathbf{p}\sigma}) &=& \frac{1}{Z^{q}} {\sum\limits_{\{n_{\mathbf{p}\sigma}\}}}^{\prime} G\{n_{\mathbf{p}\sigma}\} \nonumber \\ &\times&  \left[1-(1-q)\frac{\sum\limits_{\mathbf{p},\sigma} n_{\mathbf{p}\sigma} (\varepsilon_{\mathbf{p}}-\mu)}{T}\right]^{\frac{q}{1-q}},
\end{eqnarray}
where the prime symbol denotes the total sum without the summation over $n_{\mathbf{p}\sigma}$. It should be stressed that the exact results in the integral representation for the mean occupation numbers (see Eqs.~(\ref{10}) and (\ref{13})) and the partition function (\ref{11}) can be found in Refs.~\cite{Parvan2017a,Parvan19}. In this paper, we are rather interested in the factorization approximation, calculations for which are given in the following sections. 

\subsection{Factorization approximation}
Let us consider the factorization approximation adopted by B\"{u}y\"{u}kkili\c{c} {\it et al.} in Ref.~\cite{Buyukkilic}, which implies the following replacement:
\begin{eqnarray}\label{15}
   && \left[1-(1-q)\frac{\sum\limits_{\mathbf{p},\sigma} n_{\mathbf{p}\sigma} (\varepsilon_{\mathbf{p}}-\mu)}{T}\right]^{\xi} \nonumber \\  && \qquad \simeq \prod\limits_{\mathbf{p},\sigma}
    \left[1-(1-q)\frac{ n_{\mathbf{p}\sigma} (\varepsilon_{\mathbf{p}}-\mu)}{T}\right]^{\xi},
\end{eqnarray}
where $\xi$ is a real parameter. Substituting Eq.~(\ref{15}) into Eq.~(\ref{11}) and considering $\xi=1/(1-q)$, we obtain
\begin{eqnarray}\label{17}
  Z &=& \sum\limits_{\{n_{\mathbf{p}\sigma}\}}  G\{n_{\mathbf{p}\sigma}\} \prod\limits_{\mathbf{p},\sigma}  \left[1-(1-q)\frac{n_{\mathbf{p}\sigma} (\varepsilon_{\mathbf{p}}-\mu)}{T}\right]^{\frac{1}{1-q}} \nonumber \\
  &=& \prod_{\mathbf{p},\sigma} \sum\limits_{n_{\mathbf{p}\sigma}=0}^{K}  g(n_{\mathbf{p}\sigma})\left[1-(1-q)\frac{n_{\mathbf{p}\sigma} (\varepsilon_{\mathbf{p}}-\mu)}{T}\right]^{\frac{1}{1-q}}, \;\;\;\;\;
\end{eqnarray}
where $g(n_{\mathbf{p}\sigma})=1$ for the Fermi-Dirac and Bose-Einstein statistics of particles, and $g(n_{\mathbf{p}\sigma})=1/n_{\mathbf{p}\sigma}!$ for the Maxwell-Boltzmann statistics of particles. Now substituting Eq.~(\ref{15}) into Eq.~(\ref{10}) or Eqs.~(\ref{12})--(\ref{14}) and considering $\xi=q/(1-q)$, we get
\begin{eqnarray}\label{16a}
 && \langle n_{\mathbf{p}\sigma} \rangle = \frac{1}{Z^{q}} \sum\limits_{\{n_{\mathbf{p}\sigma}\}} n_{\mathbf{p}\sigma} G\{n_{\mathbf{p}\sigma}\} \nonumber \\ &\times&  \prod\limits_{\mathbf{p},\sigma} \left[1-(1-q)\frac{ n_{\mathbf{p}\sigma} (\varepsilon_{\mathbf{p}}-\mu)}{T}\right]^{\frac{q}{1-q}} \nonumber \\
  &=& \frac{1}{Z^{q}} \sum\limits_{n_{\mathbf{p}\sigma}=0}^{K} n_{\mathbf{p}\sigma} g(n_{\mathbf{p}\sigma})\left[1-(1-q)\frac{n_{\mathbf{p}\sigma} (\varepsilon_{\mathbf{p}}-\mu)}{T}\right]^{\frac{q}{1-q}}  \nonumber \\ &\times&
  {\prod\limits_{\mathbf{p},\sigma}}^{\prime} \sum\limits_{n_{\mathbf{p}\sigma}=0}^{K} g(n_{\mathbf{p}\sigma})\left[1-(1-q)\frac{n_{\mathbf{p}\sigma} (\varepsilon_{\mathbf{p}}-\mu)}{T}\right]^{\frac{q}{1-q}}, \;\;\;\;
\end{eqnarray}
where the prime symbol denotes the product of all the states except $\mathbf{p},\sigma$. Using Eq.~(\ref{17}), we have
\begin{widetext}
\begin{equation}\label{16}
  \langle n_{\mathbf{p}\sigma} \rangle = \frac{\sum\limits_{n_{\mathbf{p}\sigma}=0}^{K} n_{\mathbf{p}\sigma} g(n_{\mathbf{p}\sigma})\left[1-(1-q)\frac{n_{\mathbf{p}\sigma} (\varepsilon_{\mathbf{p}}-\mu)}{T}\right]^{\frac{q}{1-q}} }{\sum\limits_{n_{\mathbf{p}\sigma}=0}^{K} g(n_{\mathbf{p}\sigma})\left[1-(1-q)\frac{n_{\mathbf{p}\sigma} (\varepsilon_{\mathbf{p}}-\mu)}{T}\right]^{\frac{q}{1-q}}}
  \prod_{\mathbf{p},\sigma} \left[ \frac{\sum\limits_{n_{\mathbf{p}\sigma}=0}^{K} g(n_{\mathbf{p}\sigma})\left[1-(1-q)\frac{n_{\mathbf{p}\sigma} (\varepsilon_{\mathbf{p}}-\mu)}{T}\right]^{\frac{q}{1-q}}}{\left(\sum\limits_{n_{\mathbf{p}\sigma}=0}^{K}  g(n_{\mathbf{p}\sigma})\left[1-(1-q)\frac{n_{\mathbf{p}\sigma} (\varepsilon_{\mathbf{p}}-\mu)}{T}\right]^{\frac{1}{1-q}}\right)^{q}}\right].
\end{equation}
\end{widetext}

In the case of the Fermi-Dirac statistics of particles $(K=1)$, we obtain the following expressions for the single-particle distribution function and the partition function in the Tsallis unnormalized statistics in the factorization approximation:
\begin{eqnarray}\label{18}
  \langle n_{\mathbf{p}\sigma} \rangle &=& \frac{\chi}{\left[1-(1-q)\frac{\varepsilon_{\mathbf{p}}-\mu}{T}\right]^{\frac{q}{q-1}}+1}, \\ \label{18a}
  \chi &=& \prod\limits_{\mathbf{p},\sigma} \left[ \frac{1+\left[1-(1-q)\frac{\varepsilon_{\mathbf{p}}-\mu}{T}\right]^{\frac{q}{1-q}}}{\left(1+\left[1-(1-q)\frac{\varepsilon_{\mathbf{p}}-\mu}{T}\right]^{\frac{1}{1-q}}\right)^{q}}\right]
\end{eqnarray}
and
\begin{equation}\label{19}
  Z = \prod_{\mathbf{p},\sigma} \left(1+\left[1-(1-q)\frac{\varepsilon_{\mathbf{p}}-\mu}{T}\right]^{\frac{1}{1-q}}\right).
\end{equation}
Comparing Eq.~(\ref{18}) with Eq.~(\ref{ii}), we observe that the single-particle distribution function for the Fermi-Dirac statistics of particles in the Tsallis unnormalized statistics in the factorization approximation differs from the same distribution function obtained in Ref.~\cite{Buyukkilic} by the factor $\chi$, which is not equal to one in general, and by power $q$ in the quantity $\left[1-(1-q)\frac{\varepsilon_{\mathbf{p}}-\mu}{T}\right]^{\frac{q}{q-1}}$ in the denominator of Eq.~(\ref{18}).

\subsection{Additional factorization approximation}
To evaluate the summation in Eq.~(\ref{16}) for the Bose-Einstein and Maxwell-Boltzmann statistics of particles we use an `additional factorization approximation':
\begin{equation}\label{20}
    \left[1-(1-q)\frac{ n_{\mathbf{p}\sigma} (\varepsilon_{\mathbf{p}}-\mu)}{T}\right]^{\xi} \simeq \left[1-(1-q)\frac{\varepsilon_{\mathbf{p}}-\mu}{T}\right]^{\xi n_{\mathbf{p}\sigma}},
\end{equation}
where $\xi$ is a real parameter.

But before that, we would like to mention that a general discussion on the factorization approximation \'{a} la B\"{u}y\"{u}kkili\c{c} {\it et al.} \cite{Buyukkilic} as well as the `additional factorization' approximation depicted in Eq.~(\ref{20}) is presented in Appendix A. In addition to that, we also discuss the connection between the `additional factorization' approximation and the factorization approximation used by Hasegawa in Ref. \cite{Hasegawa} in this appendix.

Substituting Eq.~(\ref{20}) into Eq.~(\ref{16}), we obtain the mean number of particles and the partition function for the Bose-Einstein statistics of particles $(K=\infty)$ as
\begin{eqnarray}\label{21}
  \langle n_{\mathbf{p}\sigma} \rangle &=& \frac{\chi}{\left[1-(1-q)\frac{\varepsilon_{\mathbf{p}}-\mu}{T}\right]^{\frac{q}{q-1}}-1}, \\ \label{21a}
  \chi &=& \prod\limits_{\mathbf{p},\sigma} \left[\frac{1-\left[1-(1-q)\frac{\varepsilon_{\mathbf{p}}-\mu}{T}\right]^{\frac{q}{1-q}}}{\left(1-\left[1-(1-q)\frac{\varepsilon_{\mathbf{p}}-\mu}{T}\right]^{\frac{1}{1-q}}\right)^{q}}\right]^{-1}
\end{eqnarray}
and
\begin{equation}\label{22}
  Z = \prod_{\mathbf{p},\sigma} \left(1-\left[1-(1-q)\frac{\varepsilon_{\mathbf{p}}-\mu}{T}\right]^{\frac{1}{1-q}}\right)^{-1}.
\end{equation}
For the Maxwell-Boltzmann statistics of particles $(K=\infty)$, we have
\begin{eqnarray}\label{23}
  \langle n_{\mathbf{p}\sigma} \rangle &=& \chi \left[1-(1-q)\frac{\varepsilon_{\mathbf{p}}-\mu}{T}\right]^{\frac{q}{1-q}}, \\ \label{23a}
   \chi &=& \frac{e^{\sum\limits_{\mathbf{p},\sigma} \left[1-(1-q)\frac{\varepsilon_{\mathbf{p}}-\mu}{T}\right]^{\frac{q}{1-q}}}}{\left(e^{\sum\limits_{\mathbf{p},\sigma} \left[1-(1-q)\frac{\varepsilon_{\mathbf{p}}-\mu}{T}\right]^{\frac{1}{1-q}}}\right)^{q}}
\end{eqnarray}
and
\begin{equation}\label{24}
  Z = e^{\sum\limits_{\mathbf{p},\sigma} \left[1-(1-q)\frac{\varepsilon_{\mathbf{p}}-\mu}{T}\right]^{\frac{1}{1-q}}}.
\end{equation}
It should be stressed that Eqs.~(\ref{18})--(\ref{19}) and (\ref{21})--(\ref{24}) can be rewritten in a general form as
\begin{eqnarray}\label{25}
  \langle n_{\mathbf{p}\sigma} \rangle &=& \frac{\chi}{\left[1-(1-q)\frac{\varepsilon_{\mathbf{p}}-\mu}{T}\right]^{\frac{q}{q-1}}+\eta}, \\ \label{25a}
  \chi &=& \prod\limits_{\mathbf{p},\sigma} \left[ \frac{1+\eta\left[1-(1-q)\frac{\varepsilon_{\mathbf{p}}-\mu}{T}\right]^{\frac{q}{1-q}}}{\left(1+\eta\left[1-(1-q)\frac{\varepsilon_{\mathbf{p}}-\mu}{T}\right]^{\frac{1}{1-q}}\right)^{q}}\right]^{\frac{1}{\eta}}
\end{eqnarray}
and
\begin{equation}\label{26}
  Z = \prod_{\mathbf{p},\sigma} \left(1+\eta\left[1-(1-q)\frac{\varepsilon_{\mathbf{p}}-\mu}{T}\right]^{\frac{1}{1-q}}\right)^{\frac{1}{\eta}},
\end{equation}
where $\eta=1$ for the Fermi-Dirac statistics, $\eta=-1$ for the Bose-Einstein statistics, and $\eta=0$ for the Maxwell-Boltzmann statistics of particles. Equations~(\ref{21}) and (\ref{23}) differ from Eqs.~(\ref{ii}) and (\ref{i}), respectively, by the factor $\chi$, which is not equal to one, and by power $q$ in the quantity $\left[1-(1-q)\frac{\varepsilon_{\mathbf{p}}-\mu}{T}\right]^{\frac{q}{q-1}}$ in the denominator of Eq.~(\ref{25}). Thus, the single-particle distribution functions for the Bose-Einstein and Maxwell-Boltzmann statistics of particles in the Tsallis unnormalized statistics in the factorization approximation are different from the corresponding distribution functions obtained in Ref.~\cite{Buyukkilic}.

\section{Single-particle distribution function of Tsallis unnormalized statistics derived in Ref.~\cite{Buyukkilic}}\label{sec4}
\subsection{Exact results}
In this section, we derive the mean occupation numbers for the Bose-Einstein, Fermi-Dirac and Maxwell-Boltzmann statistics of particles in the framework of the Tsallis unnormalized statistics in the factorization approximation using the definition of the expectation values as in Ref.~\cite{Buyukkilic}. In Ref.~\cite{Buyukkilic}, the probability distribution of microstates in Eq.~(\ref{7}) is obtained by using the definition of the generalized expectation values given by Eq.~(\ref{2}), but the single particle distribution functions are derived from the probability distribution of microstates of the Tsallis-2 statistics, given by Eq.~(\ref{7}), by using the standard expectation values of the Tsallis-1 statistics:
\begin{equation}\label{27a}
   \langle A \rangle = \sum\limits_{i} p_{i} A_{i}.
\end{equation}
Ref.~\cite{Buyukkilic} calculates the mean occupation numbers utilizing Eq.~(\ref{27a}), a definition which is given in the Tsallis-1 statistics by using the probability distribution (\ref{12}) of the Tsallis-2 statistics. They are written as
\begin{eqnarray}\label{27}
  \langle n_{\mathbf{p}\sigma} \rangle &=& \sum\limits_{\{n_{\mathbf{p}\sigma}\}} n_{\mathbf{p}\sigma} G\{n_{\mathbf{p}\sigma}\}  p\{n_{\mathbf{p}\sigma}\} \nonumber \\ &=& \sum\limits_{n_{\mathbf{p}\sigma}=0}^{K} n_{\mathbf{p}\sigma} f(n_{\mathbf{p}\sigma})
\end{eqnarray}
and
\begin{eqnarray}\label{28}
   f(n_{\mathbf{p}\sigma}) &=& \frac{1}{Z} {\sum\limits_{\{n_{\mathbf{p}\sigma}\}}}^{\prime} G\{n_{\mathbf{p}\sigma}\}  \nonumber \\ &\times& \left[1-(1-q)\frac{\sum\limits_{\mathbf{p},\sigma} n_{\mathbf{p}\sigma} (\varepsilon_{\mathbf{p}}-\mu)}{T}\right]^{\frac{1}{1-q}},
\end{eqnarray}
where the prime symbol denotes the total sum without the summation over $n_{\mathbf{p}\sigma}$ and $Z$ is given by Eq.~(\ref{11}). Thus the definition of the mean occupation number in Eqs.~(\ref{27}), (\ref{28}) is inconsistent with the Tsallis-$2$ framework.

\subsection{Factorization approximation}
With the help of Eq.~(\ref{27}), we now calculate the mean occupation numbers in the factorization approximation for different particle statistics.

Using the approximation given by Eqs.~(\ref{15}) in Eq.~(\ref{11}) and considering $\xi=1/(1-q)$, we obtain Eq.~(\ref{17}). Substituting Eq.~(\ref{15}) into Eqs.~(\ref{27}), (\ref{28}) and considering $\xi=1/(1-q)$, we get
\begin{eqnarray}\label{28a}
 && \langle n_{\mathbf{p}\sigma} \rangle = \frac{1}{Z} \sum\limits_{\{n_{\mathbf{p}\sigma}\}} n_{\mathbf{p}\sigma} G\{n_{\mathbf{p}\sigma}\} \nonumber \\ &\times&  \prod\limits_{\mathbf{p},\sigma} \left[1-(1-q)\frac{ n_{\mathbf{p}\sigma} (\varepsilon_{\mathbf{p}}-\mu)}{T}\right]^{\frac{1}{1-q}} \nonumber \\
  &=& \frac{1}{Z} \sum\limits_{n_{\mathbf{p}\sigma}=0}^{K} n_{\mathbf{p}\sigma} g(n_{\mathbf{p}\sigma})\left[1-(1-q)\frac{n_{\mathbf{p}\sigma} (\varepsilon_{\mathbf{p}}-\mu)}{T}\right]^{\frac{1}{1-q}}  \nonumber \\ &\times&
  {\prod\limits_{\mathbf{p},\sigma}}^{\prime} \sum\limits_{n_{\mathbf{p}\sigma}=0}^{K} g(n_{\mathbf{p}\sigma})\left[1-(1-q)\frac{n_{\mathbf{p}\sigma} (\varepsilon_{\mathbf{p}}-\mu)}{T}\right]^{\frac{1}{1-q}}, \;\;\;\;
\end{eqnarray}
where the prime symbol denotes the total product of all the states except $\mathbf{p},\sigma$. Using Eq.~(\ref{17}), we have
\begin{equation}\label{29}
  \langle n_{\mathbf{p}\sigma} \rangle = \frac{\sum\limits_{n_{\mathbf{p}\sigma}=0}^{K} n_{\mathbf{p}\sigma} g(n_{\mathbf{p}\sigma})\left[1-(1-q)\frac{n_{\mathbf{p}\sigma} (\varepsilon_{\mathbf{p}}-\mu)}{T}\right]^{\frac{1}{1-q}} }{\sum\limits_{n_{\mathbf{p}\sigma}=0}^{K} g(n_{\mathbf{p}\sigma})\left[1-(1-q)\frac{n_{\mathbf{p}\sigma} (\varepsilon_{\mathbf{p}}-\mu)}{T}\right]^{\frac{1}{1-q}}}.
\end{equation}

In the case of the Fermi-Dirac statistics of particles ($K=1$ and $g(n_{\mathbf{p}\sigma})=1$), we obtain
\begin{equation}\label{30}
  \langle n_{\mathbf{p}\sigma} \rangle = \frac{1}{\left[1-(1-q)\frac{\varepsilon_{\mathbf{p}}-\mu}{T}\right]^{\frac{1}{q-1}}+1}.
\end{equation}
The single-particle distribution function (\ref{30}) for the Fermi-Dirac statistics of particles is the same as distribution function (64) of Ref.~\cite{Buyukkilic}. However, the single-particle distribution function (\ref{30}) does not recover the distribution function (\ref{18}) of the Tsallis unnormalized statistics in the factorization approximation.

\subsection{Additional factorization approximation}
Substituting Eq.~(\ref{20}) into Eq.~(\ref{29}) and considering $\xi=1/(1-q)$ and $g(n_{\mathbf{p}\sigma})=1$, we obtain the mean occupation numbers for the Bose-Einstein statistics of particles $(K=\infty)$ as
\begin{equation}\label{31}
  \langle n_{\mathbf{p}\sigma} \rangle = \frac{1}{\left[1-(1-q)\frac{\varepsilon_{\mathbf{p}}-\mu}{T}\right]^{\frac{1}{q-1}}-1}.
\end{equation}
The single-particle distribution function (\ref{31}) for the Bose-Einstein statistics of particles is the same as the distribution function (44) of Ref.~\cite{Buyukkilic}. However, the single-particle distribution function (\ref{31}) does not recover the distribution function (\ref{21}) of the Tsallis unnormalized statistics in the factorization approximation.

Substituting Eq.~(\ref{20}) into Eq.~(\ref{29}) and considering $\xi=1/(1-q)$ and $g(n_{\mathbf{p}\sigma})=1/n_{\mathbf{p}\sigma}!$, we obtain the mean occupation numbers for the Maxwell-Boltzmann statistics of particles $(K=\infty)$ as
\begin{equation}\label{32}
  \langle n_{\mathbf{p}\sigma} \rangle = \left[1-(1-q)\frac{\varepsilon_{\mathbf{p}}-\mu}{T}\right]^{\frac{1}{1-q}}.
\end{equation}
The single-particle distribution function (\ref{32}) for the Maxwell-Boltzmann statistics of particles is the same as the distribution function (47) of Ref.~\cite{Buyukkilic}. However, the single-particle distribution function (\ref{32}) does not recover the distribution function (\ref{23}) of the Tsallis unnormalized statistics in the factorization approximation.

It should be stressed that Eqs.~(\ref{30})--(\ref{32}) can be rewritten in a general form as
\begin{equation}\label{33}
  \langle n_{\mathbf{p}\sigma} \rangle = \frac{1}{\left[1-(1-q)\frac{\varepsilon_{\mathbf{p}}-\mu}{T}\right]^{\frac{1}{q-1}}+\eta},
\end{equation}
where $\eta$ is a parameter defined below Eq.~(\ref{26}).

It is apparent from Eq.~\ref{33} that the mean occupation numbers (\ref{33}) of the Tsallis unnormalized statistics in the factorization approximation obtained in Ref.~\cite{Buyukkilic} do not coincide with the mean occupation numbers given by Eq.~(\ref{25}) of the Tsallis unnormalized statistics in the factorization approximation obtained in the present paper. This implies that the results for the classical and quantum single-particle distribution functions of the Tsallis unnormalized statistics in the factorization approximation obtained in Ref.~\cite{Buyukkilic} are not correct, and the link between the Tsallis nonextensive statistics in the factorization approximation and the phenomenological Tsallis distributions is not yet established.

\section{Summary and conclusions}\label{sec5}
To summarize, in this article we revisit the link between the Tsallis unnormalized statistical formulation and the phenomenological Tsallis distributions (Eqs.~(\ref{i}) and (\ref{ii})) established in Ref.~\cite{Buyukkilic}. By using the expression for the probability of microstates and the definition of expectation values required by the Tsallis 2 statistics, we have proven that the expressions for the single-particle distribution functions of the Maxwell-Boltzmann, Bose-Einstein and Fermi-Dirac statistics (Eqs.~(\ref{i}) and (\ref{ii})) obtained in Ref.~\cite{Buyukkilic} are not connected with the Tsallis unnormalized statistics. We have pointed out that in Ref.~\cite{Buyukkilic}, the probability of microstates was taken from the Tsallis-2 statistics, while the definition of the expectation value used for the calculation of the mean occupation number, was that defined in the Tsallis-1 statistics. This is inconsistent and does not lead to the results which comply with the second law of thermodynamics (maximum entropy principle). According to these observations we conclude that the connection between the phenomenological quantum and classical single particle distributions which have widely been used in literature and the Tsallis nonextensive statistics still remains to be clarified.

\vskip0.2in
{\bf Acknowledgments:} This work was supported in part by the joint research project of JINR and IFIN-HH. We are indebted to D.V.~Anghel and S.~Grigoryan for stimulating discussions.

\vskip0.05in
\appendix
\section{Factorization approximation}
\subsection{B\"{u}y\"{u}kkili\c{c} {\it et al.} factorization}
In its general form, the B\"{u}y\"{u}kkili\c{c} {\it et al.} factorization may be summarized in the following way:
\begin{eqnarray}
\label{buyukkilicapprox}
f(n_1x_1+n_2x_2+\ldots n_sx_s) 
\approx f(n_1x_1) f(n_2x_2)\ldots f(n_sx_s), \nonumber\\
\end{eqnarray}
where in our case, $n_j$ are integer numbers, $x_j\in\mathbb{R},~\forall j$, and $f(0)=1$.
\subsection{Additional factorization}
\label{subsec2}
The additional factorization approximation used in Eq.~(\ref{20}) boils down to using the following replacement for any (say the $j$-th) term at the right hand side of (\ref{buyukkilicapprox}):
\begin{eqnarray}
\label{additionalapprox}
f(n_jx_j) &\approx& f(x_j) f(x_j)\ldots f(x_j) ~(n_j~\mathrm{times})\nonumber\\
&=& (f(x_j))^{n_j}.
\end{eqnarray}
\subsection{Hasegawa factorization}
The Hasegawa factorization approximation used in Eq.~(101) of Ref.~\cite{Hasegawa} can be described by the following recursive replacement:
\begin{eqnarray}
\label{additionalapprox1}
f\big( (n_j+1)x_j \big) &\approx& f(n_jx_j) f(x_j)~~,
\end{eqnarray}
which is equivalent to the `additional factorization approximation' described in \ref{subsec2}.

\end{document}